\documentstyle[epsfig]{aipproc}

\begin{document}

\title{Theory and Observations of Type I X-Ray Bursts from
Neutron Stars}
 
\author{Lars Bildsten} 
\address{Institute for Theoretical Physics and Department of Physics\\
Kohn Hall, University of California, Santa Barbara, CA 93106}

\lefthead{Thermonuclear Burning on Accreting Neutron Stars}
\righthead{Lars Bildsten}
\maketitle

\begin{abstract}
I review our understanding of the thermonuclear instabilities on
accreting neutron stars that produce Type I X-Ray bursts. I emphasize
those observational and theoretical aspects that should interest the
broad audience of this meeting. The easily accessible timescales of
the bursts (durations of tens of seconds and recurrence times of hours
to days) allow for a very stringent comparison to theory.  The largest
discrepancy (which was found with {\it EXOSAT} observations) is the
accretion rate dependence of the Type I burst properties. Bursts
become less frequent and energetic as the global accretion rate ($\dot
M$) increases, just the opposite of what the spherical theory
predicts.  I present a resolution of this issue by taking seriously
the observed dependence of the burning area on $\dot M$, which implies
that as $\dot M$ {\it increases}, the accretion rate per unit area
{\it decreases}. This resurrects the unsolved problem of knowing where
the freshly accreted material accumulates on the star, equally
relevant to the likely signs of rotation during the bursts summarized
by Swank at this meeting. I close by highlighting the Type I bursts
from GS~1826-238 that were found with {\it BeppoSAX} and {\it RXTE}.
Their energetics, recurrence times and temporal profiles clearly indicate
that hydrogen is being burned during these bursts, most likely by the
rapid-proton (rp) process.
\end{abstract}

\section{Introduction} 

The gravitational energy release from matter accreted onto a neutron
star (NS) of mass $M$ and radius $R$ is $GMm_p/R\approx \ 200 \ {\rm
MeV}$ per nucleon, much larger than that released from thermonuclear
fusion ($E_{nuc}\approx 5$ MeV per nucleon when a solar mix goes to
iron group elements). Hansen and Van Horn (1975) showed that the
burning of the accumulated material in the NS atmosphere occurred in
radially thin shells and so was susceptible to a thermal
instability. Evidence of the instability came soon after with the
discovery of recurrent Type I X-ray bursts from low accretion rate
($\dot M <10^{-9} \ M_\odot \ {\rm yr^{-1}}$) NSs. The successful
association of the thermal instabilities found by Hansen \& Van Horn
(1975) with the X-ray bursts made a nice picture of a recurrent cycle
that consists of fuel accumulation for several hours followed by a
thermonuclear runaway that burns the fuel in $\sim 10-100$ seconds (see
Lewin, van Paradijs and Taam 1995 for an overview and references).
The observational quantity $\alpha$ (defined as the ratio of the
time-averaged accretion luminosity to the time-averaged burst
luminosity) is close to the value expected
(i.e. $\alpha=(GM/R)/E_{nuc}\approx 40$) for a thermonuclear burst
origin.

Though our basic understanding from 25 years ago is unchanged, 
we now know much more about how the thermal instability depends on
$\dot M$, both theoretically and observationally. It is this
comparison that I emphasize, as it provides many important
lessons that are likely applicable to thin shell flashes on
accreting white dwarfs (classical novae); where 100-1000 yr
recurrence times prohibit such detailed comparisons. I focus
solely on NSs accreting at $\dot M> 10^{-10} \ M_\odot \ {\rm
yr^{-1}}$, which is appropriate for most persistently bright Low Mass
X-ray Binaries (in particular the ``Z'' and ``Atoll'' sources of
Hasinger \& van der Klis 1989). These NSs are weakly magnetic,
with $B<10^{10}$G.

 I start by reviewing the simplest aspects of the physics of the
accumulation and ignition of the fresh fuel on the NS (leaning heavily
on results from my Bildsten 1998 review article, to which I refer the
reader for the complete set of original references). I then discuss
the {\it EXOSAT} observations of the $\dot M$ dependence of the Type I
X-Ray burst properties and speculate that a solution to these puzzles
is possible if freshly accreted matter accumulates near the
equator. This problem, as well as the observations of nearly coherent
oscillations during the burst (summarized by Swank at this meeting)
are the first good indicators of the breaking of spherical symmetry. I
close with a detailed discussion of the Type I bursts from the binary
GS 1826-234, which is a beautiful example of limit-cycle mixed
hydrogen/helium burning.

\section{Accumulation, Ignition, Explosion}

  Once the freshly accreted hydrogen and helium has thermalized and
become part of the ``star'', it undergoes hydrostatic compression from
the new material that is continuously piled on.  The extreme gravity
on the NS surface compresses the fresh fuel to ignition densities and
temperatures within a few hours to days.\footnote{ The physics of the
compression and burning depends on the accretion rate per unit area,
$\dot m\equiv \dot M/A_{acc}$, where $A_{acc}$ is the covered area of
fresh material. I sometimes quote numbers for both $\dot m$ and $\dot
M$. When I give $\dot M$, I have assumed $A_{acc}=4\pi R^2\approx
1.2\times 10^{13} \ {\rm cm^2}$.}  The short thermal time in the
atmosphere (only $\sim 10$ s at the ignition location, $P\approx
10^{22}-10^{23} \ {\rm erg \ cm^{-3}}$) compared to the time to
accumulate the material (hours to days) makes the compression far from
adiabatic. Indeed, the temperature contrast from the photosphere to
the burning layer is a factor of ten; whereas the density contrast
exceeds $10^4$.

The temperature exceeds $10^7$ K in most of the
accumulating atmosphere, so that hydrogen burns via the CNO
cycle and we can neglect the pp cycles. 
At high temperatures ($T>8\times 10^7 \ {\rm K}$), the
timescale for proton captures becomes shorter than the subsequent
$\beta$ decay lifetimes, even for the slowest
$^{14}$N(p,$\gamma$)$^{15}$O reaction. The hydrogen then burns in the
``hot'' CNO
\begin{equation} 
^{12}{\rm C}(p,\gamma)^{13}{\rm N}(p,\gamma)^{14}{\rm
O}(\beta^+)^{14}{\rm N}(p,\gamma)^{15}{\rm O}(\beta^+)^{15} {\rm
N}(p,\alpha)^{12} {\rm C}, 
\end{equation} 
cycle and is limited to $5.8\times 10^{15} Z_{\rm CNO} {\rm \ ergs \
g^{-1} \ s^{-1}}$, where $Z_{\rm CNO}$ is the mass fraction of CNO and
is {\it independent of temperature}. The hydrogen burns this way in
the accumulating phase when $\dot m> 900 \ {\rm g \ cm^{-2} \ s^{-1}}
(Z_{CNO}/0.01)^{1/2}$ and is thermally stable.  The amount of time it
takes to burn the hydrogen is $\approx (10^3/Z_{\rm CNO}) \ {\rm s}$,
or about one day for solar metallicities. For lower $\dot m$'s, the
hydrogen burning is thermally unstable and is the trigger for the Type
I burst.

 The slow hydrogen burning during the accumulation 
allows for a unique burning regime at high $\dot m$'s. 
This simultaneous H/He burning occurs when $\dot m > (2-5)\times 10^{3}
\ {\rm g \ cm^{-2} \ s^{-1}} (Z_{CNO}/0.01)^{13/18}$, as at these high
rates the fluid element is compressed to helium ignition conditions
long before the hydrogen is completely burned (Lamb and Lamb 1978,
Taam and Picklum 1978). The strong temperature
dependence of the helium burning rate (and lack of any weak
interactions) leads to a strong thin-shell instability for
temperatures $T< 5 \times 10^8 \ {\rm K}$ and causes the Type I X-Ray
burst for these $\dot m$'s. The critical condition of thin burning
shells ($h\ll R$) is true before burning and remains so even during
the flash (when temperatures reach $10^9 \ {\rm K}$) as the large
gravitational well on the neutron star requires temperatures of order
$10^{12} \ {\rm K}$ for $h\sim R$. Stable burning sets in at higher
$\dot M$'s (comparable to the Eddington limit) when the helium burning
temperature sensitivity finally becomes weaker than the cooling rate's
sensitivity (Ayasli \& Joss 1982 and Taam, Woosley \& Lamb
1996). 

  For solar metallicities, there is a narrow window of $\dot m$'s
where the hydrogen is completely burned before the helium ignites. In
this case, a pure helium shell accumulates underneath the
hydrogen-burning shell until densities and pressures are reached for
ignition of the pure helium layer. The recurrence times of these
bursts must be longer than the time to burn all of the hydrogen, so
pure helium flashes should have recurrence times in excess of a
day and $\alpha\approx 200$. To summarize, in order of increasing
$\dot m$, the regimes of unstable burning we expect to
witness from NSs accreting at sub-Eddington rates ($\dot m<10^5 
\ {\rm g \ cm^{-2} \ s^{-1}}$) are (Fujimoto, Hanawa \& Miyaji 1981,
Fushiki and Lamb 1987):
\begin{enumerate}

\item{Mixed hydrogen and helium burning triggered by thermally 
unstable hydrogen ignition for $\dot m < 900 \ {\rm g \ cm^{-2} \
s^{-1}}$ ($\dot M< 2\times 10^{-10} M_\odot \ {\rm yr^{-1}}$).} 

\item{Pure helium shell ignition for 
$900 \ {\rm g \ cm^{-2} \ s^{-1}}< \dot m < \ (2-5)\times 10^3 
\ {\rm g \ cm^{-2} \ s^{-1}}$ following completion of hydrogen burning.}

\item{Mixed hydrogen and helium burning triggered by thermally 
unstable helium ignition for $\dot m > (2-5)\times 10^3 \ {\rm g \
cm^{-2} \ s^{-1}}$ ($\dot M>4.4-11.1\times 10^{-10} M_\odot \ {\rm
yr^{-1}}$).} 
\end{enumerate} 

\noindent 
The transition $\dot m$'s are for $Z_{CNO}\approx 0.01$. Reducing
$Z_{CNO}$ lowers the transition accretion rates and, more importantly,
makes the $\dot m$ range for pure helium ignition quite narrow.  We
now discuss what happens as the thermal instability develops into a
burst and what observational differences are to be expected between a
pure helium ignition and a mixed hydrogen/helium ignition.

  The flash occurs at fixed pressure, and the increasing temperature
eventually allows the radiation pressure to dominate. 
For an ignition column of $10^8 \ {\rm g \ cm^{-2}}$, the
pressure is $P=gy\approx 10^{22} \ {\rm ergs \ cm^{-3}}$, so
$aT_{max}^4/3\approx P$ gives a maximum temperature $T_{max}\approx
1.5\times 10^9 {\rm \ K}$. For pure helium flashes, the fuel rapidly
burns (since there are no limiting weak interactions) and the local
Eddington limit is often exceeded, leading to a radius expansion burst
and a duration set mostly by the time it takes the heat to
escape, $\sim 5-10$ seconds.

 When hydrogen and helium are both present, the high temperatures
reached during the thermal instability easily produces elements far
beyond the iron group (Hanawa et al. 1983; Wallace \& Woosley 1984;
Hanawa and Fujimoto 1984) via the rapid-proton (rp) process of Wallace
and Woosley (1981).  This burning starts a few seconds after the
initial helium flash (see Hanawa and Fujimoto 1984 for an
illuminating example) that makes new seed nuclei and increases the
temperature. The rp process burns hydrogen by successive proton captures
and $\beta$ decays.  The seed nuclei move up the proton-rich side
of the valley of stability (much like the r-process which occurs by
neutron captures on the neutron rich side) more or less limited by the
$\beta$-decay rates. Theoretical work shows that the end-point
of this time-dependent burning is at elements far heavier than iron
(Hanawa and Fujimoto 1984, 
Schatz et al. 1997, Koike et al. 1999).\footnote{In steady-state burning at
$\dot M>10^{-8} M_\odot {\rm yr^{-1}}$ Schatz et al. (1999) showed
that the rp-process burns all of the hydrogen and ends at nuclei with
$A$ near 100.  } The long series of $\beta$ decays allows for
energy release 10-100 seconds after the burst has started. We thus 
expect a mixed hydrogen/helium burst to last much longer than 
a pure helium burst.

\section{Observations of $\dot M$ Dependencies} 
  
The 3.8 day orbit of {\it EXOSAT} was an excellent match for the
long-term monitoring of the Type I bursters needed to reveal the
dependence of their nuclear burning behavior on $\dot M$.\footnote{I
hope that the equally well-matched {\it Chandra} and {\it XMM}
satellites will devote as much time to Type I bursters. As I will make
clear from this discussion, detailed spectroscopy during the bursts
would be very informative.}  While in a particular burning regime, we
expect that the time between bursts should decrease as $\dot M$
increases since it takes less time to accumulate the critical amount
of fuel at a higher $\dot M$.  Exactly the {\it opposite} behavior was
observed from many low accretion rate ($\dot M <  10^{-9} M_\odot \
{\rm yr^{-1}}$) NSs. A particularly good example is 4U 1705-44, where
the recurrence time increased by a factor of $\approx 4$ when $\dot M$
increased by a factor of $\approx 2$ (Langmeier et al. 1987, Gottwald
et al. 1989).  If the star is accreting matter with $Z_{CNO}=10^{-2}$
then these accretion rates are at the boundary between unstable helium
ignition in a hydrogen-rich environment at high $\dot M$ and unstable
pure helium ignition at lower $\dot M$. The expected change in burst
behavior as $\dot M$ increases would then be to more energetic and
more frequent bursts. This was not observed.

\begin{figure}
\centerline{\epsfig{file=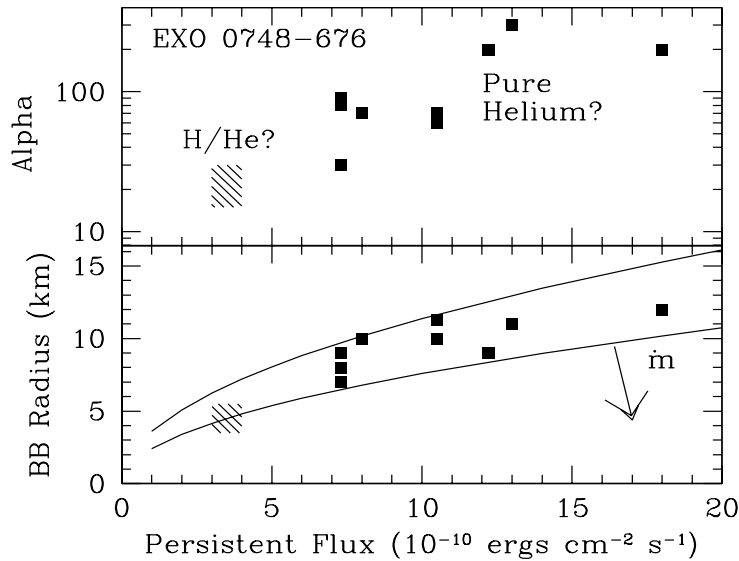,width=3.85in,height=3in}}
\vspace{3pt}
\caption{Properties of the Type I X-ray bursts from EXO 0748-676 from
Gottwald et al. (1986).  The value of $\alpha$ (top panel) and the
apparent black-body radius ($R_{app}$ for $d=10 \ {\rm kpc}$)
in the tail of the burst, are
shown as a function of the persistent X-ray flux, $F_x=L_x/4\pi d^2$.
The solid squares denote those that are well measured (for some of
these, the $\alpha$ parameters and $R_{app}$ are lower limits; see
Gottwald et al. 1986).  The hatched region is where burst properties
cluster at low $F_x$.  The lower (upper) solid line on the bottom
panel denotes where the accretion rate per unit area is constant at
$\dot m=8.3\times 10^{3} \ {\rm g \ cm^{-2} \ s^{-1}}$ ($\dot
m=3.7\times 10^{3} \ {\rm g \ cm^{-2} \ s^{-1}}$). The arrow points in
the direction of increasing $\dot m$.
\label{areafig}}
\vspace{-15pt}
\end{figure}

Other NSs showed similar behavior.  van Paradijs, Penninx \& Lewin
(1988) tabulated this effect for many bursters and concluded that
increasing amounts of fuel are consumed in a less visible way than
Type I X-ray bursts as $\dot M$ increases.  The following trends were
always found as $\dot M$ increases:
\begin{itemize}
\item{The recurrence time increases from 2-4 hours to $>$
  day.}
\item{The bursts  burn less of the accumulated fuel, with $\alpha$
increasing from $\approx 40$ to $>100$ (see top panel of Figure \ref{areafig}).}
\item{The duration of the bursts decrease from $\approx 30$ s to
$\sim 5$ s.}
\end{itemize}
\noindent 
The low $\dot M$ bursts look like mixed hydrogen/helium burning
(namely, energetic and of long duration from the rp-process) whereas
the high $\dot M$ bursts look like pure helium burning (not so
energetic, recurrence times typically long enough to have burned the
hydrogen to helium before the burst and short duration due to the lack
of any weak interactions). The simplest explanation would be 
to say that the NS has transitioned from the low $\dot M$ mixed
burning regime (noted as 1 in \S II) to the higher $\dot M$ pure
helium burning (noted as 2 in \S II). For this to be true, these
NSs should be accreting at $\approx 10^{-10}M_\odot \ {\rm yr^{-1}}$
in the lower $\dot M$ state and a factor of 4-5 higher in the high 
$\dot M$ state. However, these estimates are not consistent with the
observations. 

  van Paradijs et al. (1988) estimated the $\dot M$ from those
bursters which have shown Eddington limited radius expansion
bursts. For these systems, the ratio of the persistent flux to the
flux during radius expansion measures the accretion rate in units of
the Eddington accretion rate ($2\times 10^{-8} M_\odot {\rm
yr^{-1}}$). They showed that most bursters accrete at rates $\dot M
\approx (3-30)\times 10^{-10} \ M_\odot {\rm \ yr^{-1}}$, at least a
factor of three (and typically more) higher than the calculated rate
where such a transition should occur. Moreover, if the accretion rates
were as low as needed, the recurrence times for the mixed
hydrogen/helium burning would be about 30 hours, rather than the
observed 2-4 hours.  Fujimoto et al. (1987) discussed in some detail
the challenges these observations present to a spherically symmetric
model, while Bildsten (1995) attempted to resolve this by having much
of the thermally unstable burning occur via slow deflagration fronts
that do not lead to Type I bursts, but rather slow hour-long flares. 

  Another comparably embarrassing conundrum is the lack of regular
bursting from the six ``Z'' sources (Sco X-1, Cyg X-2, GX 5-1, GX
17+2, GX 340+0, GX 349+2) which are accreting at $3\times
10^{-9}-2\times 10^{-8}\
M_\odot \ {\rm yr^{-1}} $. These NSs very rarely show Type I bursts,
and when they do, they are so infrequent that the resulting $\alpha$
values are usually $>10^3$ (see Kuulkers et al. 1997 and Smale 1998
for examples and discussions).  In other words, these bursts are clearly
not responsible for burning all of the accreted fuel, whereas theory
clearly says that these objects should be burning nearly all of their
fuel unstably in the mixed hydrogen/helium regime (noted as 3
above). The same mystery holds for the Atoll sources with $\dot M\sim
10^{-9} M_\odot \ {\rm yr^{-1}}$ (GX 3+1, GX 13+1, GX 9+1 and GX 9+9),
which at best are infrequent bursters.  

\section{Accumulation in the Equator?} 

Many of these puzzles are resolved by relaxing our spherical
symmetry presumption and allowing the fresh material 
to only cover a fraction of the star prior to
igniting. There are observational hints that this is happening, as
the other clear trend (in addition to those noted in the previous
section) found by {\it EXOSAT} was an increase in the apparent
black-body radius ($R_{app}$) as $\dot M$ increased (see bottom panel
in Figure \ref{areafig}). This parameter is found by spectral fitting
in the decaying tail of the Type I bursts and is susceptible to
absolute spectral corrections (see discussion in Lewin, van Paradijs
and Taam 1995) that will hopefully be resolved with {\it XMM}
observations. In a similar vein, van der Klis et al. (1990) found that
the temperature of the burst at the moment when the flux was one-tenth
the Eddington limit decreased as $\dot M$ increased (hence a larger
area) for the Atoll source 4U~1636-53. In total, these observations
raise the distinct possibility that the covered area increases enough
with increasing $\dot M$ so that the accretion rate per unit area
actually {\it decreases}.

  By interpreting the measured $R_{app}$ as an indication of the
fraction of the star that is covered by freshly accreted fuel, we can
measure directly $\dot m=\dot M/4\pi R_{app}^2$, which is independent
of the distance to the source, as $F_x=GM\dot M/4\pi d^2 R$ gives
$\dot m\approx (F_xR/GM)(d/R_{app})^2$. The bottom panel of Figure
\ref{areafig} shows data for the burster EXO~0748-676. The lower
(upper) solid lines are curves where the local accretion rate is
constant at $\dot m=8.3\times 10^{3} \ {\rm g \ cm^{-2} \ s^{-1}}$
($\dot m=3.7\times 10^{3} \ {\rm g \ cm^{-2} \ s^{-1}}$). The arrow
points in the direction of increasing $\dot m$. The points at higher
$\dot M$ (as inferred from $F_x$) tend to lie at comparable or
slightly lower $\dot m$. The radius
increase appears adequate to offset the $\dot M$ increase.  In
addition, for this source, the inferred values of $\dot m$ are in the
range where the NS is transitioning from the mixed H/He burning at
high $\dot m$ to the pure helium case at lower $\dot m$.  More
physically stated, the data point to the possibility that, as $\dot M$
increases, the area increases fast enough to allow the hydrogen to
complete its burning before high enough pressures are reached for
helium ignition. If such small covering areas persist to the higher
$\dot M$'s of the Z sources, then their apparently stable nuclear
burning is easily explained.

We know that these NSs accrete from a disk formed in the
Roche lobe overflow of the stellar companion. However, there are still
debates about the ``final plunge'' onto the NS surface. Some advocate
that a magnetic field controls the final infall, while others prefer an
accretion disk boundary layer. This is now an important issue to
resolve, both for the reasons I have noted here as well as for the
oscillations seen during the bursts. If material is placed in the
equatorial belt, it is not clear that it will stay there very long. If
angular momentum was not an issue, the lighter
accreted fuel (relative to the ashes) would cover the whole star
quickly.  However, on these rapidly rotating neutron stars, the fresh
matter added at the equator must lose angular momentum to get to the
pole. This competition (namely understanding the spreading of a
lighter fluid on a rotating star) has just been recently investigated
by Inogamov and Sunyaev (1999), to which I refer the interested
reader.

\section{An Example of Mixed H/He Bursts} 

 Despite the complications I discussed in the previous sections, there
are times when Type I bursters behave in a near limit cycle manner,
with bursts occurring nearly periodically as $\dot m$ apparently stays
at a fixed value for a long time. The most recent (and beautiful!)
example of such a Type I burster is GS~1826-238. Ubertini et
al. (1999) show that during 2.5 years of monitoring with the {\it
BeppoSAX} Wide Field Camera, 70 bursts were detected from this object
with a quasi-periodic recurrence time of $5.76\pm 0.62$ hours. The
persistent flux during this bursting period was $F_x\approx 2\times
10^{-9} \ {\rm erg \ cm^{-2} \ s^{-1}}$ (Ubertini et al. 1999, In 't
Zand et al. 1999, Kong et al. 2000), which when combined with the
measured ratio of $R_{app}/d$ (about 9 km at 8 kpc; Kong et al. 2000)
gives $\dot m\approx 8\times 10^3 \ {\rm g \ cm^{-2} \ s^{-1}}$,
safely in the mixed H/He burning regime. 

If at the upper limit
distance of 8 kpc (in 't Zand et al. 1999) the global accretion rate is
$\dot M\approx 10^{-9} M_\odot \ {\rm yr^{-1}}$ and apparently the
whole star is covered with fresh material. No coherent pulsations have
been detected in these bursts, so we do not know this NS's rotation
rate.

\begin{figure}
\centerline{\epsfig{file=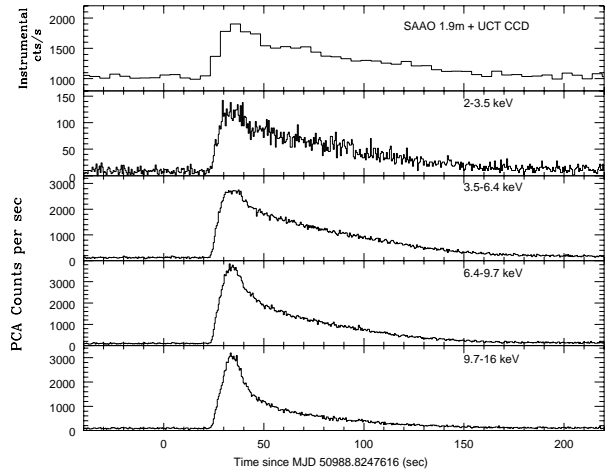,width=2.5in,height=3.25in,angle=-90}}
\vspace{3pt}
\caption{A montage showing the Type I burst profile from 
GS 1826-238 from RXTE 
as well as the reprocessed optical emission (from Kong et
al. 2000). The long duration of the burst is indicative of the delayed
energy release from the rapid proton (rp) process. 
\label{1826fig}}
\vspace{-15pt}
\end{figure}
 
 The type I bursts from this source are a ``textbook'' case for the
mixed hydrogen/helium burning expected at these accretion rates. The
estimated $\dot m$ gives an accumulated column on the NS prior to the
burst of $1.6\times 10^8 \ {\rm g \ cm^{-2}}$, just what is
expected from theory. These quasi-periodic bursts allow for a very secure
measurement of $\alpha\approx 50$, which implies a nuclear energy
release of $4 \ {\rm MeV}$ per accreted nucleon for a $1.4 M_\odot$,
10 km NS.  Energy releases this large can only come about via hydrogen
burning and the long ($>100 $s ) duration of the bursts are consistent
with the expected long-time energy release from the rp-process. Figure
\ref{1826fig} shows the time profile of such a burst seen with RXTE
(Kong et al. 2000) in a few different energy bands. Though these data
were taken to study the optical reprocessing (top panel is the
simultaneous optical burst), they provide an important confirmation of
the delayed energy release expected when hydrogen is burning via an
rp-process. The resemblence of these profiles to Hanawa and Fujimoto's
(1984) theoretical results are striking.

The upcoming launch of the {\it High Energy Transient Explorer}
should provide comparable long-term coverage as the Wide Field
Camera on {\it BeppoSAX} and gather more information on such nice 
bursts from many more LMXB's.

\section{Conclusions} 
 
 I hope I have made the case that Type I bursts from neutron stars are
still very interesting to study in their own right and provide
important lessons to those studying thin shell flashes in other
astrophysical contexts. The detailed comparison provided by the
neutron star systems is likely telling us to seriously consider the
possibility and repercussions of fuel preferentially accumulating in
the equatorial region. Another place where this might prove
immediately applicable are the recurrent novae, where currently one
infers high accretion rates and white dwarf masses in order to get the
short recurrence times of 20-50 years (Livio 1994). These constraints
are relaxed if we allow for a smaller covering fraction. I am not the
first to say this, but hopefully the Type I burst observations make
such a warning harder to ignore!
 
  Jean Swank reviewed the observations of nearly coherent oscillations
in the 300-600 Hertz range during many Type I bursts and so I will not
summarize those results here. Though I am convinced that these
modulations are intimately connected to rapid stellar rotation, there
are still important unresolved questions. The ones that bother me the
most are:
\begin{enumerate}
\item {What causes the asymmetry at late times in the burst, long after the
peak?} 
\item{Why does the modulation appear sometimes at twice the spin
frequency?}
\item{How does the burning front really spread on a rapidly
rotating star? Ignition at one spot is plausible, but we do not
understand how the ignited/hot fuel spreads around a rapidly rotating
star.}
\end{enumerate} 
It is even an open question as to why, from a particular NS, only some
bursts show these oscillations. Before any meaningful theoretical work
can be carried out, what is needed is the phenomenology of the
oscillations in the context of the well established burst
phenomenology I have discussed here. My current mental tabulations
point to a complete abscence of oscillations during the long bursts
(even during the long rise) indicative of mixed H/He burning. All
reported detections of oscillations during bursts that I am aware of
are from short duration, high $\alpha$ bursts.

  I thank Andrew Cumming, Erik Kuulkers, and Michiel van der Klis for
 discussions and comments on the manuscript. My recent plunge into the
 {\it EXOSAT} Type I burst literature occurred when I was the CHEAF
 Visiting Professor at the Astronomical Institute ``Anton Pannekoek''
 of the University of Amsterdam. I thank them for the hospitality and
 Michiel for reminding me to eventually place his 1990 article on 4U
 1636-53 in a meaningful context. This research was supported by NASA
 via grant NAG 5-8658 and by the NSF under Grant PHY 94-07194.
 L. B. is a Cottrell Scholar of the Research Corporation.

\end{document}